\documentclass{aa}

\usepackage{graphics,psfig}

\begin{document}      

\title{Observations of pulsars at 9 millimetres}

\author{
        O.~L\"ohmer \inst{1},
        A.~Jessner \inst{1},
        M.~Kramer \inst{2},
        R.~Wielebinski \inst{1}
        O.~Maron \inst{3,1}.
        }

\institute{Max-Planck-Institut f\"ur Radioastronomie, Auf dem H\"ugel 69,
              D-53121 Bonn, Germany
    \and
           University of Manchester, Jodrell Bank Centre for Astrophysics,
           Manchester M13 9PL, UK
    \and
           J.~Kepler Astronomical Centre, Pedagogical University,
           Lubuska 2, PL-65-265 Zielona Gora, Poland
           }

\offprints{M.~Kramer, e-mail: jessner@mpifr-bonn.mpg.de}
\date{Received / Accepted}

\titlerunning{Pulsar observations at 9 Millimetres}
\authorrunning{O.~L\"ohmer {\it et al.}}

%%%%%%%%%%%%%%%%%%%%%%%%%%%%%%%%%%%%%%%%%%%%%%%%%%%%%%%%%%%%%%%%%%%%%
\abstract{}{The behaviour of the pulsar spectrum at high
radio frequencies can provide decisive information about the nature of 
the radio emission mechanism.}
{We report recent observations of a selected 
   sample of pulsars at $\lambda$=9~mm (32 GHz) with the 100-m Effelsberg
  radio telescope.}
{Three pulsars, PSR B0144+59, PSR B0823+26, and
  PSR B2022+50, were detected for the first time at this
  frequency. We confirm the earlier flux density measurements for a 
  sample of six pulsars, and we are able to place upper flux density 
  limits for another 12 pulsars. We find that all pulsar spectra 
  have a simple form that can be described
  using only three parameters, one of which is the lifetime 
  of short nano-pulses in the emission region.  
  The study of the transition region from coherent to
  incoherent emission needs  further and more sensitive 
  observations at even higher radio 
  frequencies.}{}

\keywords{\it pulsars: individual (PSR B0144+59, PSR B0823+26,
PSR B2022+50) -- radiation mechanisms: nonthermal -- radio continuum:
stars}

\maketitle

%%%%%%%%%%%%%%%%%%%%%%%%%%%%%%%%%%%%%%%%%%%%%%%%%%%%%%%%%%%%%%%%%%%%%
\section{Introduction\label{intro}}

Almost 40 years after the discovery of pulsars, it is still not possible to give
a detailed explanation of the radio emission process, and a large number of
fundamental questions remain (e.g.~Lorimer \& Kramer 2005)\nocite{lk05} to be
answered.  Advances in our theoretical understanding of the radio emission and
the associated conditions in the pulsar magnetosphere
(e.g.~Petrova~2001\nocite{pet01}, Melrose et al.~2006\nocite{mmkl06}) are
still largely driven by observational progress (e.g.~Kramer et
al.~2006\nocite{klo+06}).

In recent years, most observational studies pertaining to emission processes 
have concentrated on polarisation
properties (e.g.~Karastergiou \& Johnston~2004\nocite{kj04}) or on the drift
behaviour of sub-pulses (e.g.~Weltevrede et al.~2006\nocite{wes06}). However,
as past studies of other discrete radio sources have demonstrated, the
investigation of radio frequency spectra can be crucial for identifying
of the underlying emission process (e.g. Conway et al. 1963\nocite{ckl63}).
Although we may expect that similar detailed studies of pulsar spectra can be
helpful in solving the `pulsar problem', such studies (e.g.~Malofeev et
al.~1994)\nocite{mgj+94} have so far not been successful in establishing a
relationship between pulsar spectra and other observed properties.  It turned
out that significant correlations between spectral properties and pulsar spin
parameters could not be found or did not survive more detailed scrutiny
(Maron et al.~2000\nocite{mkkw00}).  It is therefore more likely that
spectral studies of separate profile components (\cite{kra94}) or
even of individual pulses (\cite{kkg+03,kj06}) are better suited
to addressing questions about the emission process, as the average pulse profile
seen at different frequencies is dominated by different components with
different spectral behaviours. Differences in the flux density spectra of
separate components can indeed explain certain features in observed spectra of
the {\em average} profiles, such as breaks in power laws, but may also
contribute to the observed polarisation evolution with frequency (Karastergiou
et al.~2005).  Other authors have, however, claimed that the general decrease in
polarised emission at high frequencies is due to a frequency-dependent
decrease in the coherence of pulsar radio emission (e.g.~Xilouris et
al.~1996\nocite{xkj+96}).

The steepness of pulsar spectra (mean spectral index of $\alpha =
-1.6$ with indices between $\alpha \sim 0$ and $\alpha \sim -3$ for power
law spectra, $S\propto \nu^{\alpha}$, Maron et al.~2000\nocite{mkkw00}) 
may be caused
by the mechanisms that enforce the coherence of pulsar radiation as has 
been suggested by \cite{mic91}.  
This concept is based on a model in which an initially incoherent radiation process 
produces a coherent output by means of antenna mechanisms or resonances. 
The radiation produced by other considered processes, especially
maser mechanisms and plasma radiation, is however inherently coherent.
Should the first class of models be applicable, then pulsar
radiation would be a combination of coherent and incoherent emission, where
the coherent emission is seen to dominate at radio frequencies,
establishing a typical coherence length $\lambda_c$ beyond which
incoherent emission should dominate. Interestingly, the non-thermal
emission of pulsars observed at infrared and optical frequencies is
indeed believed to be due to incoherent synchrotron emission 
(\cite{cl02c}).

The study of pulsars at very high radio frequencies, where the coherent
emission process might become less dominant, is an obvious procedure
for investigating a possible intrinsic coherence length and the contribution of
an additional incoherent emission component.  So far, the flux density
spectrum of pulsar emission in the transition region between the coherent
radio emission and the incoherent infrared emission is unknown. 
We therefore aim to investigate the exact position and nature of this 
transition from coherent to incoherent emission through access to the spectral
region where the spectral index is flattening or perhaps showing an upturn.

The information gap between high radio
and infrared frequencies was first pointed out after early observations of the
Crab pulsar (e.g. Smith 1977\nocite{smi77}) at radio and optical wavelengths.
The Crab pulsar, among a few others, also provides direct evidence about the fine
details of the radio-emission process in the form of giant radio pulses (GRPs).
These have been observed up to 15 GHz for the Crab pulsar (Hankins private communication,
Kondratiev et al. in preparation) and occur during all rotation phases where the average
pulse profile shows radio emission (\cite{js05}). It is worthwhile investigating 
the possibility of the spectrum, or parts of it, being the natural result of a superposition
of a very large number of very short pulses, similar to the  `nano-shots' (\cite{eh06}) 
observed for the Crab pulsar. One can show that a simple form of the spectrum will be the result, 
depending -- apart from a scale factor -- on at most two more parameters, one of which is the
typical decay time of the nano-shots.

The filling of the spectral gap, hence the attempt to learn more about the radio
emission process by studying this crucial frequency range, was one of the
motivations for the technically challenging observations of pulsars at
mm-wavelengths (Wielebinski et al.~1993\nocite{wjkg93}, Kramer et al.~1996).
Only at high frequencies will propagation effects become negligible
for the majority of pulsars, their intrinsic emission properties may reveal 
themselves only here.  
This is commonly appreciated for the interstellar medium, but it may even be
important for propagation effects in the pulsar magnetosphere
(e.g.~Hoensbroech \& Lesch 1999\nocite{hl99}).

As we summarise in the following, the first observations of pulsars at
mm-wavelengths provided some unexpected results, which triggered
further follow-up studies.  The results of such a study are described in this paper, 
confirming the results for the known detected pulsars but with an additional 
enlargement of the numbers of observed pulsars to provide better
statistics.

In this paper we briefly review the main
properties of pulsar spectra, before we describe new observations of
pulsars at mm-wavelengths and present their results. We then compare
these new results with previous measurements before we discuss the
implications of our results and our conclusions are drawn.

\section{Pulsar spectra}

Prior to the observations by Wielebinski et al.~(1993\nocite{wjkg93}),
it was commonly believed that the generally steep pulsar spectrum
continues to the highest radio frequencies (e.g.~Manchester \& Taylor
1977\nocite{mt77}, Malofeev et al.~1994\nocite{mgj+94}). On the other 
hand, one can even
find claims in the literature that pulsar spectra show a ``cut-off''
at about 10 GHz or so (e.g. \cite{gbi93}).  This
erroneous impression was caused by the lack of large sensitive telescopes
and high-frequency receivers at that time. 
The absence of any cut-off was demonstrated by the first
detection of four pulsars at 33.8~GHz ($\lambda=$ 8.9mm \cite{wjkg93}).
 In fact, two
pulsars were stronger at these very high radio frequencies than was
previously thought. In contrast to predictions based on the
extrapolation of the spectra measured up to $\sim 24$~GHz, the
measured flux densities of these two sources (PSRs B1929+10 and
B2021+51) were factors of five to ten higher than expected. This
apparent spectral turn-up was further investigated by Kramer et
al.~(1996\nocite{kxj+96}). With their observations between $\sim 27$~GHz and
$\sim 35$~GHz, they confirmed the results for B1929+10 and B2021+51,
but also established a normal continuation of the known spectrum for six
additional sources.  The four pulsars strongest at 32~GHz were
later also successfully observed at 43~GHz ($\lambda \sim 7$mm),
underlining the impression that pulsar spectra show
unusual behaviour at high radio frequencies; i.e.~they may become flatter
or even possibly show a turn-up at mm-wavelengths (Kramer et
al.~1997\nocite{kjdw97}). The first and so far only
detection\footnote{ Here we are not counting the recent detection of radio emission at 144 GHz 
(\cite{FC07b}) from the magnetar 
XTE J1809-194 as typical {\it pulsar} emission. Although it is accepted that these
sources are  rotating and magnetised neutron stars, their variability and spectra differ
significantly from ordinary pulsars (\cite{ksj+07}, K. Lazaridis et al. in preparation, B. Stappers et al. in preparation). 
An implicit assumption that their emission process is the same as that of ordinary pulsars
cannot be justified at present. }
 of a radio pulsar at 87 GHz ($\lambda= 3$mm) with the 30-m Pico
Veleta telescope (Morris et al.~1997\nocite{mkt+97}) shows a
continuation of the relatively flat spectrum of the detected pulsar
B0355+54.
However, recent reports suggest that turn-ups in pulsar
spectra may even occur at much lower radio frequencies (\cite{km04}). 
Clearly, more studies of more pulsars
at mm-wavelengths are required.

%%%%%%%%%%%%%%%%%%%%%%%%%%%%%%%%%%%%%%%%%%%%%%%%%%%%%%%%%%%%%%%%%%%%%
\section{Observations\label{obs}}

We selected a sample of 15 pulsars from the catalogue of
Maron et al.\ (2000)\nocite{mkk+00}, from the list of brightest
pulsars, and from the list of pulsars with unusual frequency evolution
given by von Hoensbroech \& Lesch (1999). All observations were made
with the 100-m Effelsberg radio telescope of the Max-Planck-Institut
f\"ur Radioastronomie in Bonn, Germany. We observed the sources at
three separate epochs with good to excellent weather conditions: 2002
March, 2002 September, and 2003 January. Integration times of the
pulsars were between 30 and 180~min, depending on the expected flux
density and visibility of the sources.

We used an HEMT receiver installed in the secondary focus with a fixed
central frequency of 32~GHz ($\lambda= 9.37$mm). The total bandwidth of
the system was 2~GHz, with a receiver noise temperature of about
60~K. The telescope gain at this frequency is $\sim 1$K/Jy.
The receiver provided LHC and RHC signals that were digitised
and independently sampled every $P/1024$ s and synchronously folded
with the topocentric pulse period $P$, using the Effelsberg Pulsar
Observation System (EPOS, Jessner 1996\nocite{jes96}). Records of 15~s
sub-integrations were transferred to disk for off-line data analysis.

For flux-calibration purposes, we made use of an internal noise diode
fed directly  into the waveguide following the feed horn. 
EPOS allows us to switch this noise diode synchronously with the
pulse period, occupying the first 50 phase bins of the pulse period in
the measured profile. To check that the actual pulsar signal,
expected to be relatively weak at 32 GHz, does not overlap with the 
strong, artificial calibration signal, we also observed each pulsar briefly at 
4.85 GHz and 8.35 GHz, making sure that it is positioned correctly, and then
switched to the high frequency receiver in less than
 30s that way preserving the pulse phase for all frequencies. The
strong and clearly visible pulsar signals observed at the two lower 
frequencies could then be directly related to the possible signal
of the high frequency emission, greatly facilitating confirmation of
a possible detection.

%
%%%%%%%%%%%%%%%%%%%%%%%%%%%% FIG.1 %%%%%%%%%%%%%%%%%%%%%%%%%%%%%%%%%%
\begin{figure*}
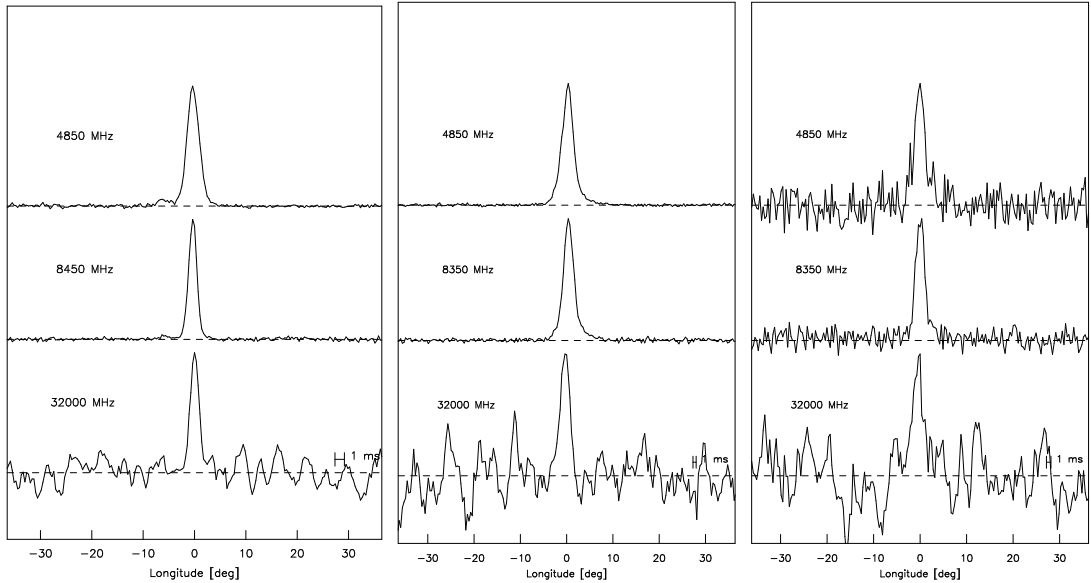

\centering
\begin{tabular}{c@{\hspace*{0.2cm}}c@{\hspace*{0.2cm}}c}
\mbox{\psfig{file=6806fig1.ps,width=5cm}} &
\mbox{\psfig{file=6806fig2.ps,width=4.5cm}} &
\mbox{\psfig{file=6806fig3.ps,width=4.5cm}}
\end{tabular}
\caption{\label{fig:profiles} Time-aligned profiles of the three PSRs
B0144+59, B0823+26 and B2022+50 at centre frequencies of 4.85, 8.35
and 32~GHz}
\end{figure*}

Flux densities were obtained for all three frequencies by comparing the
pulse profiles to the corresponding signal of the noise diode, which
itself was calibrated by observing known reference sources during
regular pointing observations. The continuum references 
(NGC 7027, 3C 286, and NRAO 150) were observed approximately 
every hour, checking both pointing and focus stability. Typical
pointing errors were RMS$\sim 5''$, compared to a beamwidth of 23''
at 32 GHz. Flux densities for the chosen reference sources were
obtained from the catalogue of Peng et al.~(2000).\nocite{pkkw00}
Uncertainties in the resulting pulsar flux densities are estimated
to be about 20\% for a single observation.

%%%%%%%%%%%%%%%%%%%%%%%%%%%% Table 1 %%%%%%%%%%%%%%%%%%%%%%%%%%%%%%%%%%
\begin{table}
\begin{center}
\caption{Results of observations at 32~GHz\label{tab:results}}
\vspace{0.1cm}
\begin{tabular}{lrrrl}
\hline\hline\noalign{\smallskip}
\multicolumn{1}{c}{PSR} & \multicolumn{1}{c}{Pulses} &
\multicolumn{1}{c}{Time} & Flux Density & no. of\\
 &\multicolumn{1}{c}{total} & \multicolumn{1}{c}{(min)} & (mJy) & 
measurements\\
\hline
\noalign{\smallskip}
\multicolumn{5}{c}{\it New Detections:}\\
B0144+59 & 30552 & 99.8 & $0.062\pm 0.006$ & 2\\
B0823+26 & 24752 & 218.9 & $0.023\pm 0.0010$ & 2 \\
         & 12516 & 111.0 & $<0.170$ & 0 (K95)$^*$ \\
B2022+50 & 9720 & 60.4 & $0.046\pm 0.009$ & 1\\
\noalign{\smallskip}
\multicolumn{5}{c}{\it Re-detections:}\\
B0355+54 & 17480 & 45.6 & $0.76 \pm 0.14$ & 2\\
         & 57095 & 148.8 & $0.8 \pm 0.2$ & 6 (K95) \\
B1133+16 & 2148  & 42.5 & $0.055 \pm 0.06$ & 1\\
         & 13920 & 275.4 & $0.03 \pm 0.02$ & 2  (K95) \\
B1706$-$16 & 8712 & 94.8 & $0.07 \pm 0.01$ & 1\\
           & 11154 & 121.2 & $0.06\pm 0.01$ & 2 (K95) \\
B1929+10 & 15708 & 59.3 & $0.19 \pm 0.02$ & 1\\
         & 133188 & 502.8 & $0.21\pm 0.01$ & 6 (K95) \\
B2020+28 & 6278 & 35.9 & $0.06 \pm 0.01$ & 1\\
         & 41065 & 235.2 & $0.09\pm 0.02$ & 1 (K95) \\
B2021+51 & 3752 & 33.1 & $0.28 \pm 0.03$ & 1\\
         & 61096 & 349.8 & $0.323\pm 0.007$ & 9 (K95) \\
\noalign{\smallskip}
\multicolumn{5}{c}{\it Upper flux limits:}\\
B0154+61 & 4578 & 179.0 & $< 0.06$ & \\
B0611+22 & 32076 & 179.0 & $< 0.09$& \\
B0628$-$28 & 2256 & 46.8 & $< 0.3$& \\
B0740$-$28 & 16554 & 46.0 & $< 0.17$& \\
B1604$-$00 & 5460 & 38.4 & $< 0.13$& \\
B1642$-$03 & 7334 & 47.4 & $< 0.3$& \\
B1822$-$09 & 2888 & 37.0 & $< 0.13$& \\
B1935+25 & 6734 & 22.6 & $< 0.9$& \\
B2000+32 & 8946 & 103.9 & $< 0.08$& \\
B2319+60 & 3168 & 119.0 & $< 0.3$& \\
B2323+63 & 2490 & 59.6 & $< 0.6$& \\
B2334+61 & 4830 & 39.9 & $< 0.4$& \\
\hline
\end{tabular}
\end{center}
{$^*$ The designation (K95) refers to earlier observations reported by Kramer (1995).\nocite{kra95} }
\end{table}
%%%%%%%%%%%%%%%%%%%%%%%%%%%%%%%%%%%%%%%%%%%%%%%%%%%%%%%%%%%%%%%%%%%%%

%%%%%%%%%%%%%%%%%%%%%%%%%%%%%%%%%%%%%%%%%%%%%%%%%%%%%%%%%%%%%%%%%%%%%
\section{Results\label{RESULTS}}

\begin{figure*}
\centering
\begin{tabular}{c@{\hspace*{0.2cm}}c@{\hspace*{0.2cm}}c}
\mbox{\psfig{file=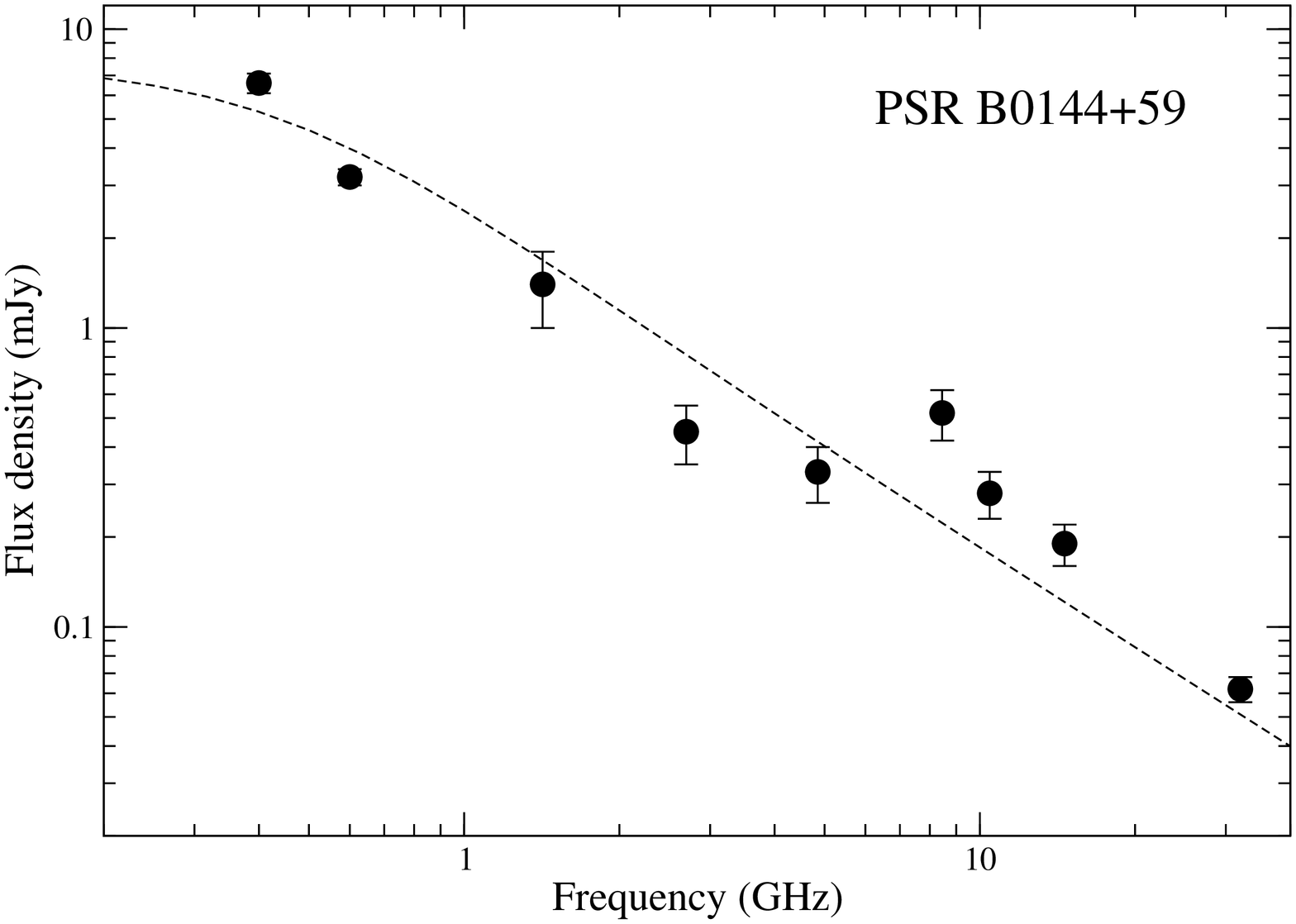,angle=-90,width=5.5cm}} &
\mbox{\psfig{file=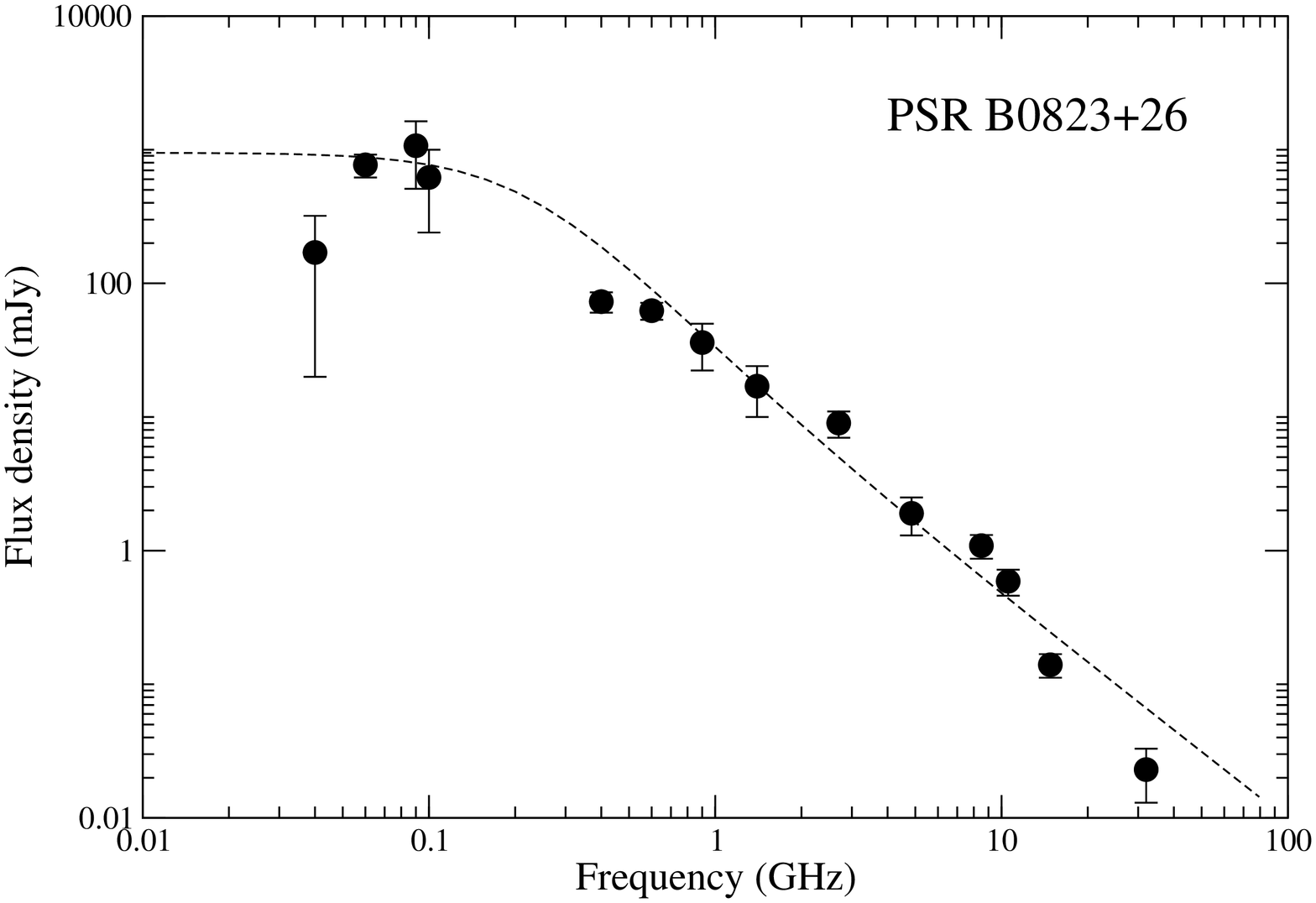,angle=-90,width=5.5cm}} &
\mbox{\psfig{file=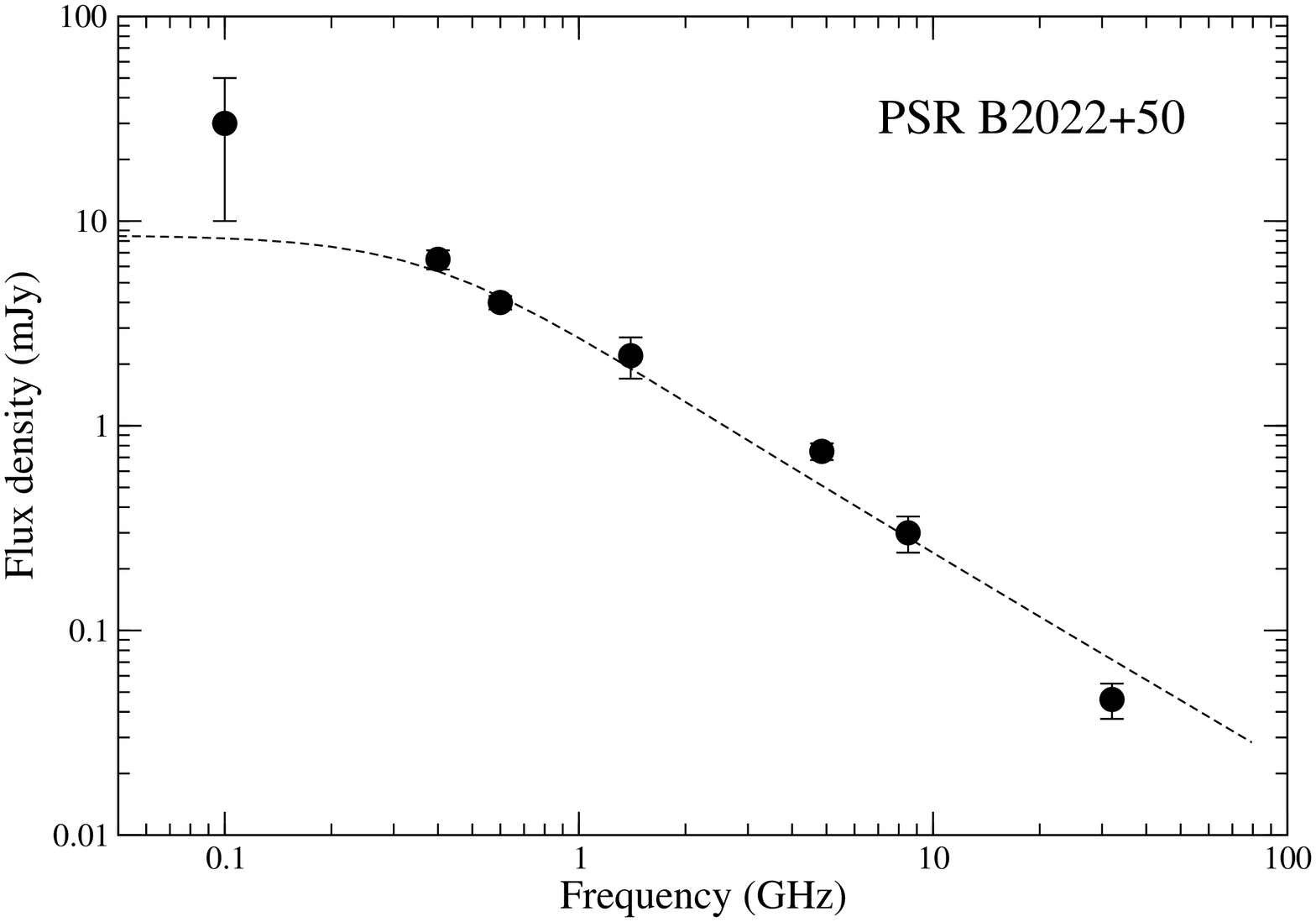,angle=-90,width=5.5cm}}
\end{tabular}
\caption{\label{fig:spectra} Resulting flux density spectra for PSRs
B0144+59, B0823+26, and B2022+50. Low-frequency data have been taken from
available literature (see text). The dashed line shows the model fit with
parameters listed in Table 2.}
\end{figure*}

In total, we observed a sample of 21 pulsars.  The results of the measurements
are summarised in Table~\ref{tab:results}.  We were able to detect PSRs
B0144+59, B0823+26, and B2022+50 at 32~GHz for the first time, determining their
flux densities at their so far highest observation frequency.  The
time-aligned profiles of these sources measured at 4.85, 8.35, and 32~GHz are
shown in Figure~\ref{fig:profiles}, while their resulting spectra are displayed in
Figure~\ref{fig:spectra}. Flux densities shown for lower frequencies are taken
from the literature (Maron et al.~2000) or from unpublished observations.

%%%%%%%%%%%%%%%%%%%%%%%%%%%% FIG.2 %%%%%%%%%%%%%%%%%%%%%%%%%%%%%%%%%%

We also successfully observed all those pulsars that were already detected
by Wielebinski et al.~(1993\nocite{wjkg93}), Kramer (1995\nocite{kra95}), and
Kramer et al.~(1997\nocite{kjdw97}). The measured flux densities
are in very good agreement with previous observations.

For those pulsars that were not detected, we estimate upper flux limits
following the procedure described by Sieber \& Wielebinski (1987\nocite{sw87}).
The upper limit is based on an estimate of five times the RMS of the noise
signal multiplied by the equivalent pulse width. For our sample we used the
pulse width $W_{50}$ at the highest known frequency as given in the EPN pulsar
data base \footnote{\tt \scriptsize
  http://www.jb.man.ac.uk/research/pulsar/Resources/epn/} or by Kramer et al.\ 
(1994\nocite{kwj+94}) or Kijak et al.\ (1998\nocite{kkwj98}).

%%%%%%%%%%%%%%%%%%%%%%%%%%%%%%%%%%%%%%%%%%%%%%%%%%%%%%%%%%%
\section{Discussion\label{discus}}
\subsection{Turn-up at high frequencies?}

Our results confirm all previously published measurements of flux densities of
pulsars at mm-wavelengths. We also present spectral information for three
newly detected pulsars at mm-wavelengths.  No new turn-up in the spectrum was
found. The spectra of the newly detected pulsars follow the trends determined
from observations at lower radio frequencies. While on one hand this gives us
great confidence in the reliability of the adopted calibration procedure, it
does suggest, on the other, that a spectral turn-up at a frequency as low
as 30 GHz is the exception rather than the rule.  We note that the spectrum of
PSR B0144+59 shows a peculiar `kink' at 3 GHz to 10 GHz, which may be
misinterpreted as a `turn-up' or definite `flattening' unless data above 10
GHz are considered (cf.~Kijak \& Maron 2004)\nocite{km04}.  We believe that this is
caused by the peculiar high-frequency profile evolution of this source, which will be
studied elsewhere. In any case, it
adds to the notion that some pulsar spectra may not be adequately described by
simple power laws.  Moreover, in view of the evidence that individual pulsar
spectra are quite different (differing not only in shape but also e.g.~in the
location of the low-frequency turn-over, spectral index, and the possible
existence of a spectral steepening (Maron et al.~2000), it would be rather
naive to assume that all pulsars show identical behaviour in the narrow
frequency range probed by our mm-observations. Hence we should not expect to
detect a spectral turn-up for {\it all} pulsars at 32 GHz.  Instead, the known
diversity of pulsar properties may indeed eventually reveal a turn-up or
flattening of pulsar spectra at higher or even at lower frequencies (Kijak \&
Maron 2004).  \nocite{km04}

\subsection{The shape of pulsar spectra}  

Although pulsar spectra are traditionally described by a power law $S_\nu
\propto S^{\alpha}$ with $\alpha$ about -1.6 (Sieber 1973\nocite{sie73}),
many of the observed spectra turned out to fit badly to that concept. 
A number of additional characteristic features are traditionally employed to describe
more complex pulsar spectra : some of them
exhibit a turn-over at low $\nu$, some have a `broken power law', some a
turn-up or flattening, etc.  The original rationale was the analogy of
synchrotron radiation, which has a power-law spectrum, so that radio astronomers
initially tried to model most observations in terms of power laws.  But one of
the implicit assumptions was that one wanted to model a temporarily continuous
and spatially large-scale radiation mechanism.  The observations have, however, come up
with strong variability on timescales from days down to nanoseconds (GRPs).
A few attempts have been made to fit another  function to a pulsar
spectrum, most notable Ochelkov \& Usov (1984), \nocite{ou84}
who proposed a six-parameter model equation: 
\begin{equation}
S(\nu)={S_0\cdot \nu^a \over \left(1 + \left({\nu\over \nu_b
      }\right)^b\right) \left(1 + \left({\nu \over \nu_c}\right)^c\right) } \  .
\end{equation}

Nearly all observed spectra can be described with a suitable choice of these
six parameters.  In Ochelkov \& Usov's model of curvature radiation of plasma
bunches, the parameters $a$, $b$, and $c$ have fixed algebraic relationships,
and only two of them should suffice to describe any spectrum. Malofeev et al.
(1994) \nocite{mgj+94} showed, however, that the model parameters were not
constrained by the already available data; hence, one may well ask oneself if
that is the  simplest possible description of a pulsar spectrum.

Do we know of other natural processes, involving  flowing charges, that
produce spectra of the observed form?  Indeed we do, solid-state currents in
everyday materials exhibit a non-thermal low frequency noise, often called
`flicker noise', `1/f-noise' or simply lf-noise. The spectrum is  very
similar to a pulsar spectrum, but at  frequencies lower by a factor of
$10^7-10^8$. That in itself is not surprising, as the conditions on the pulsar
surface and in the magnetosphere are also many orders of magnitude different
 from our experience with everyday materials.  People had already been studying 
that kind of noise in
solids and on surface boundaries in the 1950's (Pfeifer
  1959, Bess 1953)\nocite{pfe59,bes53}. It is indeed surprising how far one
can get using this very simple one-parameter theory.

We already know from observations that pulsar radio emission varies on very short
timescales. Very high-resolution observations of giant radio
pulses have so far not found a shortest timescale for the emission, although
they have reached nanosecond to sub nanosecond resolution (Hankins et 
al.~2003\nocite{hkwe03}
also Jessner et al., 2005 Eilek\& Hankins 2006)\nocite{js05} \nocite{eh06} 
for the Crab pulsar.  Eilek and Hankins (2006)\nocite{eh06} 
report an apparent increase in micro-burst duration
$\propto \nu^{-2}$, nano-shots with $\delta t < 1 $ns, and they infer that each
giant pulse is a coherent superposition of a number of powerful nano-pulses.
Based on these theoretical studies and on the observational evidence, we assume
that all pulsar radio emission is a superposition of short-lived 
($\tau \sim \rm ns$) elementary emission processes. The observed shape of the
spectrum may then be used to provide constraints on the timescales of the elementary
processes.

A shot noise model for the emission due to short-lived non-Gaussian elementary 
processes in ordinary pulsars has been proposed by Jenet et al.~(2001)\nocite{JAP2001}. 
However, in this study we use a different elementary process, 
one that has been commonly used to
model flicker noise in solids and vacuum discharges, and evaluate the expected 
observable broad band radio spectrum.

Let us assume that we
have a stochastic formation of radio emission centres that are characterised
by a localised potential $\Phi$ with a lifetime (observer system) of i.e.
$\tau_e \sim 10^{-10}$s, very similar to the nano pulses proposed i.e. by
Weatherall (1998) \nocite{wea98} and also evident in the
PIC-simulations of
Jaroschek and Lesch (2006)\nocite{jl06}.  The decay of the electric
potential of a nano-pulse may be described by $\Phi(t)=\Phi_0\cdot
e^{-t/\tau_e}$ with ($t>0$).    A
random superposition of short pulses has the simple spectrum
\begin{equation}
S(\omega)={S_0 \over 1+\omega^2\tau_e^2} \  ,
\end{equation}
and most of the observed pulsar spectra fit extremely well within a reasonable
range of $\tau_e$.  The model spectrum depends only the reference flux and a
characteristic time $\tau_e$ for the nano-burst decay time.  
Fig.~3 shows the
fit to the spectrum of B2021+51, which extends up to 43~GHz. Here we find
$\tau_e=0.09$~ns.  Although the fits using such a simple function are already
surprisingly good, they can be improved by extending the analogy to current
noise processes (Bess 1953) a bit further. 
The theory was found to describe
noise process currents through solid state boundaries quite well (Pfeifer
1959).\nocite{pfe59} Details of the derivation and its connection to a statistical
approach to pulsar radio emission processes will be given in a forthcoming
paper (Jessner et al., in preparation).  Let us assume that these nano--pulses
have a spatial cross section of $\sigma_c$ and exist only in small regions of
mean radius $r_0$ and thickness $\delta$. They are supposed to diffuse through
the region on a timescale of $\tau_0$. Then again, such a nano--pulse may be
described by the evolution of its potential $\Phi(t)=\Phi_0\left({\tau_0\over
    \tau+\tau_0}\right)^n \cdot e^{-t/\tau_e}$ with the exponent
$n={\sigma_c\tau_0 c \over \pi \delta r_0^2 }$. The spectrum from a
superposition of such pulses can again be found from their Fourier transform
and be written in the form
\begin{equation}
$$S(\omega)=S_0\left({{1+\omega^2\tau_e^{2} }\over \tau_e^2}\right)^{n-1}\cdot e^{-i(n-1)\cdot atan(\omega\tau_e)}$$.
\end{equation}
All observed pulsar radio spectra can easily be fitted using only the three
parameters $S_0$, $\tau_e$, and $n$ as seen in in the example of B2021+51 in
Fig.~3.  
%%%%%%%%%%%%%%%%%%%%%%%%%%%%%%% FIG. 3 %%%%%%%%%%%%%%%%%%%%%%%%%%%%%%%%%%%%%%
\begin{figure}
\centering
\psfig{file=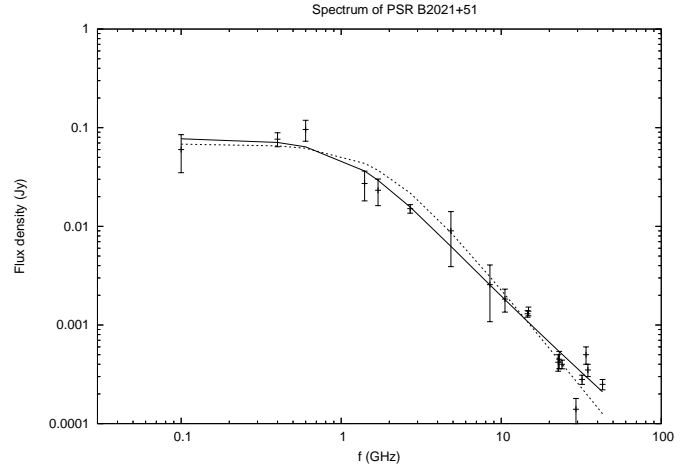,angle=-90,width=9.0cm}
\caption{Spectral fits for PSR B2021+51 using 
additional data from O. Maron et al.. 
Solid: fit with a spectrum according to Eqn.(3)
 with $S_0=1.61$Jy, $\tau_e=0.124$ns, and $n=0.275$.
Dashes: Fit with $S_\nu(\nu)={S_0 \over 1+(2\pi\nu)^2\tau_e^2}$,
here $S_0=68$~mJy and $\tau_e=0.09$ns.}
\end{figure}
We now find a larger $\tau_e=0.124$~ns, as equation (3) usually
improves the fit for the low frequency ($< 1$~GHz) part of the spectrum.
However, only the characteristic time $\tau_e$ of the nano-pulses can be
directly inferred from the fit to the spectrum, while other physical
parameters are being constrained in combination by the value of the exponent
$n$. Table 2 lists the results for those pulsars that
have been observed up to very high frequencies. The spectra obtained for the
pulsars newly detected at 32 GHz are shown in Figure 2. One notices that the
characteristic timescales $\tau_e$ are all found to be in the appropriate
range of $0.1 {\rm ~ns } < \tau_e < 2.0 {\rm ~ns}$. 
The example of GRP  sources like the Crab pulsar provides direct evidence 
that radio emission processes occur
on very similar timescales. The spectra of the above sources can be interpreted as 
indirect evidence that these are also intrinsic timescales for the emission of ordinary
pulsars, where observational constraints do not allow us to resolve the emission
on these timescales. But although the Crab pulsar does provide the best {\it direct} evidence 
for nano-pulse emission in all its components (\cite{js05}), it cannot serve as a good example
for the {\it indirect} evidence through its spectrum.  The individual
components show very distinct evolution with frequency and 
different nano-pulse characteristics(\cite{eh06}), whereas the radio fluxes are usually
 averages over all components. Furthermore, the duration of the bursts of nano-shots
scales as $\nu^{-2}$, which will add another complication to attempts to model the
radio spectrum of that unique source.

%%%%%%%%%%%%%%%%%%%%%%%%%%%% Table 2 %%%%%%%%%%%%%%%%%%%%%%%%%%%%%%%%%%
\begin{table}
\begin{center}
\caption{Parameters of spectral fits for a number of pulsars\label{tab:results2}
}
\vspace{0.1cm}
\begin{tabular}{lrrl}
\hline\hline\noalign{\smallskip}
\multicolumn{1}{c}{PSR} & \multicolumn{1}{c}{$S_0$} &
\multicolumn{1}{c}{$\tau_e$ } & n  \\
 & {(Jy)} & \multicolumn{1}{c}{(ns)}  \\
\hline
B0144+59 & 0.03 & 0.34  & 0.454 \\
B0329+59 & 22.0  & 0.16 & 0.073 \\
B0355+54 & 0.27  & 0.27  & 0.498 \\
B0628-28 & 3.09 & 0.62   & 0.114 \\
B0823+26 & 1.77  & 0.71  & 0.167\\
B0950+08 & 7.68 & 0.42  & 0.115 \\
B1133+16 & 4.41  & 0.51 & 0.103 \\
B1706-16 & 0.40 & 0.91 &  0.272 \\
B1929+10 & 2.59 & 0.57 &  0.233 \\
B2021+51 & 1.61 & 0.12 & 0.275 \\
B2020+28 & 1.54 & 0.24 & 0.196 \\
B2022+50 & 0.03 & 1.77 & 0.479 \\
\hline
\end{tabular}
\end{center}
\end{table}
%%%%%%%%%%%%%%%%%%%%%%%%%%%%%%%%%%%%%%%%%%%%%%%%%%%%%%%%%%%%%%%%%%%%%

\section{Conclusions\label{conclusion}}

We have presented the first detection of three more pulsars at
mm-wavelengths, while presenting upper limits of flux densities for 12 other
pulsars. None of these upper limits is yet low enough to establish
the nature of the flux density spectrum of those sources, the continuation of the previous 
spectral trends, or a possible flattening or turn-up at high frequencies.
 Indeed, while the three newly detected pulsars show a
continuation of the lower frequency spectra, we confirmed the flux densities
of six previously detected pulsars.  This sample of re-detected pulsars
includes PSRs B1929+10 and B2021+51, thus giving further credibility to their
published flux density measurements, which led to the suggestion of a spectral
turn-up at mm-wavelengths for these pulsars.  Establishing such a trend could
provide a possible link between the spectrum of the coherent radio emission
and the incoherent higher frequency radiation, but more sensitive observations at 
even higher
frequencies are needed to substantiate any conclusion in this direction. 

We have also found evidence that a heuristic model of pulsar radio emission, 
consisting of a superposition of a large number of short pulses of only 
nano-second duration describes observed pulsar radio spectra 
quite naturally. Apart from intensity scaling, a model fit needs only two parameters, 
one being the typical nano pulse life time $\tau_e$, the other ($n$) 
related to the geometry of the individual emission process. These processes
already play a role, with much greater magnitude for the individual nano-pulse,
 in known giant pulse emission of several pulsars.   Different emission
heights for different frequencies would have to be due to propagation effects (refraction)
in the pulsar magnetosphere. A detailed appraisal of these issues is in preparation.     

To achieve final reliable proof of these findings, it is important to
obtain even higher radio frequency data of sufficient sensitivity. Future observations
of single pulses at very high frequencies combined with very high sensitivity should 
reveal the elementary emission processes we propose here in ordinary
pulsars.  
Our present observations extend the spectra of observed pulsars into the long mm 
wavelength range. 
So far only the detection of one object, PSR 0355+54, has been made at 87 GHz with 
the Pico Veleta 30-m telescope (Morris et al. 1997) at the $S = 0.5 \pm 0.2 \rm mJy$ 
level. Observations of nine pulsars at $\lambda = 0.87 mm$  (345 GHz)  have not been 
successful (L\"ohmer et al. 2004).  The advent of ALMA should change this. 
ALMA (www.eso.org/projects/alma/) will have observational capabilities in 
ten frequency bands. Initially the frequency bands 84-116 GHz, 275-373 GHz, and 
602-720 GHz will be available. An IF bandwidth (in 2 polarisations) of  8 GHz 
will be provided.  
The flux sensitivity of ALMA is expected to be sub-mJy (about 0.1 mJy) range 
after 10 minutes of integration. Considering a 4-hour integration, a detection level 
of roughly 0.05 mJy should be reached.  
This would allow the study of the flux densities of numerous pulsars even if they 
do not show a turn-up in their spectra. By these observations the gap between radio 
and optical (IR) flux detections could be closed, giving us new information about 
the emission mechanism inherent to pulsars.
For a proper assessment of the range of coherent emission from pulsars one needs 
more information about the spectral
behaviour of pulsars at frequencies below 100 MHz where observational data is 
comparatively rare. Here pulsars are 
usually fairly strong radio sources with flux densities in the range of several Jy. 
The new generation of large low-frequency interferometers (LOFAR and possibly LWA) 
will be able to  make detailed spectral measurements
over several octaves on metre and dekametric wavelengths (Eilek et~al. 2006, 
Stappers et al.~2006). The knowledge about the upper 
and lower limits of coherent pulsar radio emission is expected 
to become a decisive constraint on pulsar radio emission theories.

%%%%%%%%%%%%%%%%%%%%%%%%%%%%%%%%%%%%%%%%%%%%%%%%%%%%%%%%%%%%%%%%%%%%%
\begin{acknowledgements}
The authors would like to thank the staff of the radio--observatory Effelsberg 
for their help and support during the observations. We would also like to thank 
Harald Lesch, Jan Kuijpers and Francis Graham-Smith for encouraging discussions 
about the shape of pulsar spectra.
We are grateful to the anonymous referee for helpful advice. 
\end{acknowledgements}

%%%%%%%%%%%%%%%%%%%%%%%%%%%%%%%%%%%%%%%%%%%%%%%%%%%%%%%%%%%%%%%%%%%%%
%\bibliographystyle{apj}
%\bibliography{journals,modrefs,psrrefs,crossrefs}

\end{document}